\documentclass[twocolumn]{webofc}
\usepackage[varg]{txfonts}   
\begin{document}
\title{On the nature of the shape coexistence and the quantum phase transition phenomena: lead region and Zr isotopes}
%
%

\author{\firstname{Jos\'e-Enrique} \lastname{Garc\'ia-Ramos}\inst{1,2}\fnsep\thanks{\email{enrique.ramos@dfaie.uhu.es}} \and
        \firstname{Kris} \lastname{Heyde}\inst{3}\fnsep\thanks{\email{kris.heyde@ugent.be}}
}

\institute{Departamento de  Ciencias Integradas y CEAFMC
  Universidad de Huelva, 21071 Huelva   
  \and
  Instituto Carlos I de F\'{\i}sica Te\'orica y Computacional, Universidad de Granada, Fuentenueva s/n, 18071 Granada, Spain   
  \and
  Department of Physics and Astronomy, Ghent University, Proeftuinstraat, 86 B-9000 Gent, Belgium
          }

\abstract{
  The goal of this contribution is to analyze the connection between
  shape coexistence and quantum phase transition, two seemingly
  unrelated phenomena that share common aspects, namely, the rapid
  change in the ground state structure along an isotope chain or the
  presence of several minima at the mean-field level. To illustrate the
  similarities and differences between both phenomena, we will focus in
  the Pb region, in particular in Pt and Hg isotopes, as well as in Zr isotopes.
}

\maketitle
\section{Introduction}
\label{intro} 
The atomic nuclei can present different shapes, namely, spherical,
axially deformed, either prolate or oblate, or even triaxial. In 
general, the shape evolves in a gradual way as one passes from a 
nucleus to its neighbor, but in certain cases the change can be very
abrupt. This is the key element for the appearance of a quantum phase
transition (QPT) \cite{Cejn10}. On the other hand, in a given nucleus different
shapes can coexist in the energy spectrum, leading to the appearance of shape
coexistence \cite{heyde11}. The presence of extra configurations can be easily singled
out in the very conspicuous parabolic behaviour in the energy
systematics centered around the mid-shell. Moreover, another relevant
indicator of the presence of intruder configurations is the lowering
of the first excited $0^+$ state that presents a minimum also around the
mid-shell. 

Along this contribution we try to disentangle which are the
relationships and differences between both phenomena. To this end, we
analyze three different chains of nuclei, Pt, Hg, and Zr which
are ideal cases to see
the competition between QPT and shape
coexistence phenomena.

\section{Shape coexistence and quantum phase transition in a nutshell}
\label{sec-sc-qpt}
Shape coexistence in nuclear physics was first proposed by Morinaga in
the 1950's
and since them it has given rise to a property of atomic nuclei that appears throughout the entire nuclear landscape, especially to
those nuclei at or near shell or sub-shell closures \cite{heyde11}. Shape coexistence
presents certain distinct experimental features: a U shape in the energy
systematics of certain excited bands, lowering of certain excited
$0^+$ states, a rapid change in the value of spectroscopic quadrupole
moments, and the existence of strong E0 transitions. All of them are
enhanced and present an almost symmetric behaviour with respect to the
corresponding mid-shell. The appearance of intruder configurations
reflects the competition between the energy gap, that
tends to maintain spherical shapes, and the residual interaction, that
favours the deformation of the nucleus and lower, in some cases
considerably, the excitation energy of the intruder states \cite{heyde11}. 

Shape coexistence can be understood in terms of two major theoretical
approaches, namely the spherical shell model and the mean
field. According to the shell model approach, the nucleus is described
accordingly as an inert core and a set of valence nucleons that occupy
certain orbits and interact amongst them through a residual two-body
interaction.
The promotion of pairs of nucleons across the shell gaps at shell and subshell closures
is the mechanism responsible of the creation of intruder configurations.
From the point of view of mean-field theory, using the self-consistent
Hartree-Fock-Bogoliubov (HFB) theory,
one obtains an energy surface depending on
certain deformation parameters. In this framework, the ground band will
correspond to the deepest minimum, but the others can be interpreted
as intruder configurations, that will appear higher in energy and 
corresponding with larger deformations.

The concept of QPT refers to the sudden change of the atomic nucleus
ground state structure as a function of a control parameter. Such
control parameter can be, e.g., the neutron number and,
therefore, a QPT can appear in an isotopic chain where the ground
state deformation varies in an abrupt way when passing from an isotope
to its neighbor.
Taking into account
that the two-neutron separation energy (S$_{2n}$) is somehow proportional to the
derivative of the energy, a first-order QPT will involve a
discontinuity in S$_{2n}$ while a second-order one a discontinuity in
the derivative of S$_{2n}$. In this contribution we are mainly
interested in the case of first-order QPT because it allows the
coexistence of two phases around the phase transition region.

An ideal framework to deal with QPTs are the algebraic models, such as
the interacting boson model (IBM) \cite{iach87}. In such a framework a
QPT can be 
modeled in terms of a Hamiltonian that is a combination of two given
symmetries of the system, combined through a control parameter, $H=x\
H_{\mbox{sym1}}+(1-x)\ H_{\mbox{sym2}}$. The onset of a QPT is denoted
by the existence of a critical value of the control parameter, x$_c$,
for which the structure of the system passes from one phase with
{\it symmetry 1} to another phase with {\it symmetry 2.}
\begin{figure}[h]
\centering
\includegraphics[width=8cm,clip]{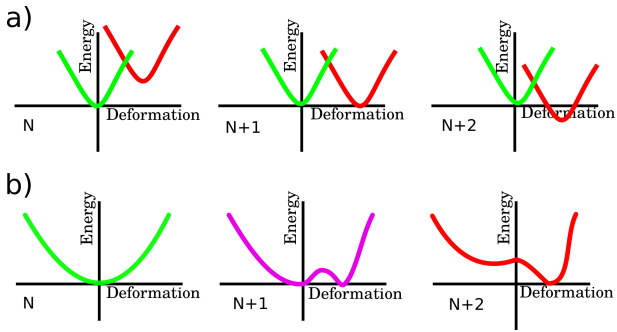}
\caption{Schematic representation (both scales in arbitrary units) of
  the  evolution of the energy surface for shape coexistence (a) and
  QPT (b) cases, as a function of the neutron number, $N$.}
\label{fig-minima-sch}  
\end{figure}

Are shape coexistence and QPT related in any way? On one hand, both
phenomena involve a rapid change in the structure of a certain set of states,
either ground or excited states. On the other hand, in both cases, at
the mean-field level, several minima coexist. In a pictorial way, in
Fig.~\ref{fig-minima-sch} we depict the evolution of two minima as a
function of the neutron number. In panel a) we treat the shape
coexistence situation where the relative position of the regular
(green) and intruder (red) minima changes as a function of the neutron
number. In this case, we assume that intruder and regular minima become
degenerated and even that the intruder describes the ground state of the
system, although
this situation is not general and in many cases the intruder band never
becomes the ground state.  In panel b) we consider the QPT case, for which
in the left most panel only one spherical minimum exists, in the
central one an energy surface with two minima develops and
finally in the right most panel the spherical minimum disappears and
only the deformed one remains. The case where the two minima coexist
leads to the existence of a QPT. Obviously both situations
present clear similarities, namely, around the region of interest two
minima coexist and the value of the deformation changes abruptly in
the central panels. However, at the quantum level one expects to observe
clear differences. For example, the height of the barrier between both
configurations should be a
clear source of differences because the states of the bands with corresponding
different wells will effectively interact for low barriers while
remain unperturbed for high ones. In the figure, we depicted the case
of a high barrier with no interacting configurations for the case of
shape coexistence, while a low one for QPT. Note that other situations
can also exist.

\section{Zr, Pt, and Hg}
\label{sec-nuclei}
The Pb region is, probably, the best example of
shape coexistence along the mass table and, indeed, in Pb up to three different
configurations have been identified. 
In the Hg and Pt nuclei, which are the nuclei we are interested in, two kind of configurations show up with a clear presence of low lying $0^+$ states, combined with a parabolic shape in the excitation energy systematics in Hg, whereas not so obvious for Pt.
On the other hand, Zr is known to exhibit the
faster change in deformation ever observed, with $^{98}$Zr almost spherical and
$^{100}$Zr a very well deformed nucleus.
\begin{figure}[h]
\centering
\includegraphics[width=8.5cm,clip]{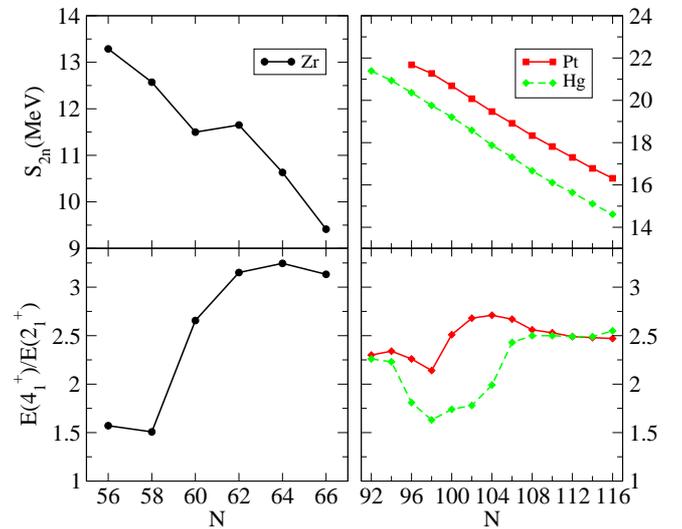}
\caption{S$_{2n}$ and $E(4_1^+)/E(2_1^+)$ as a function of the neutron
  number,$N$, for Zr,
Pt and Hg.}
\label{fig-ratio}
\end{figure}

From the point of view of a QPT, there are two key observables that can
be used as indicators of the onset of a QPT. The first one is the
S$_{2n}$ that will experience a discontinuity at the first-order transition point, and the second one is the ratio between the
excitation energies of the states $2_1^+$ and $4_1^+$,
$E(4_1^+)/E(2_1^+)$, which is related with the deformation or, in
other words, with the value of the order parameter of the QPT. In
Fig.~\ref{fig-ratio} we present the systematics of  S$_{2n}$ and
$E(4_1^+)/E(2_1^+)$, as well, for Zr and Pt and Hg. Regarding the value of
S$_{2n}$,  one notices a clear discontinuity in Zr, but a fully
linear behavior in Hg and Pt. On the other hand, $E(4_1^+)/E(2_1^+)$
shows for Zr the typical rapid change that is observed in a QPT,
passing from a value close to $2$ (even $1.5$ for $^{96-98}$Zr) to another close to $3.3$. However, in
Pt and Hg the interpretation of the systematics is not so obvious. While in Pt it
partially resembles what is observed in a QPT, in Hg there is a
noticeable dropping of the ratio around mid-shell. In the latter
two cases the systematics can be understood in terms of the
coexistence of two configurations that cross or approach very closely.   

\begin{figure}[h]
\centering
\includegraphics[width=8.5cm,clip]{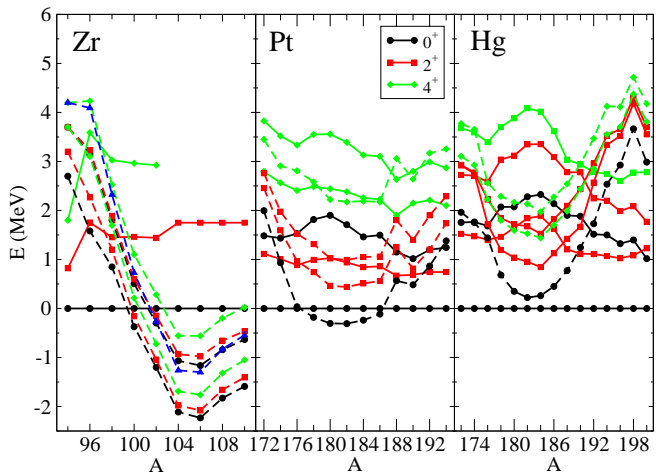}
\caption{Theoretical IBM excitation energies for selected states in Zr, Pt, and
  Hg and Hamiltonians without interactions between the regular and the
  intruder sectors, as a function of $A$. Full lines stand for regular states and dashed ones for
  intruder ones.}
\label{fig-spectrum}
\end{figure}

To shed some more light on the problem, in Fig.~\ref{fig-spectrum} we show results corresponding
to IBM calculations that were presented elsewhere (\cite{Garc09} for Pt,
\cite{Garc14} for Hg and \cite{Garc17} for Zr). In particular, the
results correspond to excitation energies of selected states, both
regular and intruder, for which the interaction term between intruder
and regular sectors of the Hamiltonian has been switched off. In these
three cases one observes a rather flat energy systematics for the
intruder states while the conspicuous parabolic behaviour for the
intruder ones centered around the mid-shell ($A=106, 182$, and $184$
for Zr, Pt, and Hg, respectively). However, several distinct features are worth to be
mentioned. First, in Zr and Pt nuclei,  a intruder
state becomes the ground state around the mid-shell, while it never happens for Hg. That means
that the structure of the ground state changes abruptly around the
mid-shell in Zr and Pt, but not in Hg. In Hg a change in the
structure 
is observed in the $2^+$ states. Note that the more realistic calculation
involves a certain degree of mixing between the regular and the
intruder states, which is rather small in Zr \cite{Garc17} and in Hg
\cite{Garc14}, but much larger 
in Pt \cite{Garc09}, which means that the observed behavior in
Pt is smoothed out in 
the more realistic calculation, but remains essentially the same in Zr
and Hg.

The above features have direct consequences in the systematics of
S$_{2n}$ and $E(4_1^+)/E(2_1^+)$. In the case of Zr, as observed
in Fig.~\ref{fig-ratio} can be easily understood in terms of the
crossing of configurations depicted in Fig.~\ref{fig-spectrum}. Indeed,
the change in the nature of the ground state produces a sudden change
in the slope of the energy systematics and therefore a discontinuity
in the value of S$_{2n}$. On the other hand, because the crossing is
so fast, not only the ground state presents an intruder character, but
also the rest of low-lying members of the yrast band involved in the energy ratio $E(4_1^+)/E(2_1^+)$, which makes that for
$A<100$ all the states present a vibrational character, while for $A\ge 100$  a rotational
one appears. On the other hand, in the case of Hg, because the ground state
exhibits a regular character and, moreover, there is no
interaction between intruder and regular sector, the linear tendency of S$_{2n}$ even at the
mid-shell can be understood. Besides, the crossing between $2_1^+$ and
$4_1^+$ 
regular and intruder states can explain the drop of the ratio
$E(4_1^+)/E(2_1^+)$ around the mid-shell. Finally, to explain the
systematics of Pt in Fig.\ref{fig-ratio} is a challenge. First, the crossing of regular and intruder $0^+$
energies would suggest a discontinuity in S$_{2n}$ as the observed one in
Zr. However experimental values show a fully linear systematics. The
reason for such a behaviour is the strong mixing between the regular
and the intruder sectors (see \cite{Garc09}). On the other hand, the energy ratio
$E(4_1^+)/E(2_1^+)$ seems to present the precursors of a QPT, but they
are hindered. Most probably, this too is due to the strong mixing
between the regular and intruder states. Note that according to
\cite{Garc14a} the value of the deformation, both in the IBM and in
the HFB calculations,  changes as it should be in a
QPT. However the strong mixing somehow hides its effect in both the S$_{2n}$ and  $E(4_1^+)/E(2_1^+)$ observables.

\section{Conclusions}
\label{sec-conclu}
We have analyzed the interplay between shape coexistence and QPT in three
chain of isotopes, namely Zr, Pt, and Hg, that seem to be the most
promising candidates to disentangle the relationship between both
phenomena. In the case of Zr, the QPT that appears around $A=100$ is
easily explained in terms of the existence of two families of states
that cross at the phase transition point. In Hg, because the ground
state never becomes of intruder character, no indications of QPT are
shown, neither for  S$_{2n}$ not for $E(4_1^+)/E(2_1^+)$. Finally, in
Pt, the crossing of regular and intruder configurations could induce a
QPT, but it is not observed in S$_{2n}$ and only partially in
$E(4_1^+)/E(2_1^+)$. The reason for that is most probably the strong
mixing between regular and intruder configurations. Therefore, we have
found a mechanism for inducing a QPT from shape coexistence, though
probably is not the only one.

Financial support from the Interuniversity Attraction Poles
Program of the Belgian State-Federal Office for Scientific,
Technical and Cultural Affairs (IAP Grant No. P7/12) and the Spanish MINECO and FEDER under Project No. FIS2014-53448-C2-2-P are acknowledged.
%

\begin{thebibliography}{}
%
%


\bibitem{Cejn10} P.\ Cejnar, J.\ Jolie, and R.F.\ Casten, Rev.\ Mod.\
  Phys.\ {\bf 82}, 2155 (2010).
\bibitem{heyde11} K.~Heyde and J.L.~Wood, Rev.\ Mod.\ Phys.\ {\bf 83}, 1467 (2011).
\bibitem{iach87}F.~Iachello and A.~Arima, {\it The Interacting Boson Model}, Cambridge University Press (1987).
\bibitem{Garc09} J.E.\ Garc\'{\i}a-Ramos and K. Heyde, Nucl.\ Phys.\ \textbf{A 825}, 39 (2009); J.E.\ Garc\'{i}a-Ramos, V. Hellemans, and K. Heyde,   Phys.\ Rev.\ C {\bf  84}, 014331 (2011).
\bibitem{Garc14} J.E.\ Garc\'{i}a-Ramos, and K. Heyde, Phys.\ Rev.\ C {\bf 89}, 014306 (2014).
\bibitem{Garc17} J.E.\ Garc\'{\i}a-Ramos and K. Heyde, work in progress.
\bibitem{Garc14a} J.E. Garc\'{i}a-Ramos, K.\ Heyde, L.M.\ Robledo, and
  R.\ Rodriguez-Guzm\'{a}n, Phys.\ Rev.\ C {\bf 89}, 034313 (2014). 
\end{thebibliography}
%
%

\end{document}